\documentstyle{aipproc}

\begin{document}
\title{Stability of Multiquark Systems \thanks{Talk presented
at the International Workshop on Hadron Physics "Effective Theories
of Low Enegy QCD", Coimbra, Portugal, September 10-15, 1999}}

\author{Fl. Stancu \thanks{e-mail
address: fstancu@ulg.ac.be.}}
\address{Institute of Physics, B. 5
University of Liege, Sart Tilman\\
B-4000 Liege 1, Belgium}

%\lefthead{LEFT head}
%\righthead{RIGHT head}
\maketitle

\begin{abstract}
%%%%%%%%%%%%%%%%%%%%%%%%%%%%%%%%%%%%%%%%%%%%%%%%%%%%%%%%%%%%%%%%%%%%
We give a brief review of developments in the field of exotic
hadrons formed of more than three quarks and/or antiquarks. In particular
we discuss the stability of multiquark systems containing heavy
flavours. We show that the gluon exchange model and the chiral
constituent quark model based
Goldstone boson (pseudoscalar meson) exchange give entirely different 
results.
%%%%%%%%%%%%%%%%%%%%%%%%%%%%%%%%%%%%%%%%%%%%%%%%%%%%%%%%%%%%%%%%%
\end{abstract}

\section*{Introduction}
%\[
%\widehat{a} + \widehat{ab} + \widehat{abc} + \widehat{abcd}
%\]
%
%%\show\frak
% 
%\[
%%      {\bf x}^{\bf x} \triangleq z 
%      {\bf x}^{\bf x}\triangleq{z} \tensor{T} \frak{E^E}=\frak{mc}^2
%%      {\bf x}^{\bf x}\triangleq {z} \tensor{T} \frak{E}=\frak{mc}^2
%\]
% 
%\[
%{\Bbb {NQRZ}} \qquad \because \eth\ggg\bigstar \therefore\blacktriangleright\rightsquigarrow \blacksquare
%\]
% 
%%%%%%%%%%%%%%%%%%%%%%%%%%%%%%%%%%%%%%%%%%%%%%%%%%%%%%%%%%%%%%%%%%%%%%%%%%%%%%%
Here we consider systems
formed of more than three quarks and/or
antiquarks ($ q^m \overline{q}^n$ with $ m+n > 3 $).
These are a category of exotic hadrons. They are "exotic" with respect 
to "ordinary" hadrons which are either mesons ($ q \overline{q}$) or
baryons ($q^3$) systems.
The existence or nonexistence of stable exotics against
strong decays is crucial for understanding some
aspects of the strong interactions.
If discovered, their properties could be an important test for
the validity of various quark models.
Most theoretical and experimental effort has been devoted 
so far to systems described by the colour
state ${\left[{222}\right]}_C$. These are the tetraquarks
$q^2\overline{q}^2$ \cite{JA77}, the pentaquarks $q^4\overline{q}$
\cite{GI87,LI87} and the hexaquarks $q^6$ \cite{JA77}.
They have a baryonic number B = 0, 1 and 2 respectively.
Multibaryon systems with B up to 9 have also been studied
within the SU(3) Skyrme model \cite{SC99}.

The most celebrated example of exotics is the H-particle (H- for hexaquark)
predicted in the context of the MIT bag model more than 20 years ago \cite
{JA77}. It is a dibaryon with the flavor content of two $\Lambda$  baryons,
i. e. it contains light quarks only. Its existence remains controversial
(see below).

From theoretical general arguments
one expects an increase in stability of
multiquark systems if they contain heavy flavours $Q = c$ or $b$.
A recent review of the experimental efforts to search for the
H-particle and charmed-strange pentaquarks can be found in \cite{ASH99}.
In summary, no evidence for the production of a deeply bound H-particle
has been observed, the production cross section being one order of
magnitude than the theoretical estimates. By deeply bound state
we understand a compact object, bound by about 80 MeV, as
in Jaffe's picture \cite{JA77}. However, a molecular type structure,
like that of the deuteron, is not excluded and there are some
suggestive signals to be confirmed.
Within the confidence level of the analyzed
experiments, no convincing evidence for the production of
pentaquarks  with the flavour content $uuds\overline{c}$ and
$udds\overline{c}$ has been observed either. However the existence
of pentaquarks is not ruled out. The analysis done so far can provide
a good starting point for future search in high statistics charm
experiments at CERN \cite{MO96} or Fermilab \cite{AI98}.

If experimentally discovered, the properties of multiquark 
systems would help to put constraints on phenomenological
interquark forces. Indeed
the theoretical predictions are model dependent. Here we are
concerned with constituent quark models which simulate the
low-energy limit of QCD and discuss theoretical predictions
for compact objects. 
We compare results from constituent quark models where
the spin-dependent term of the quark-quark interaction is described by
the chromomagnetic part of the one gluon exchange (OGE) interaction
\cite{RU75}
with results we obtained from a chiral model where the quarks
interact via Goldstone boson exchange
(GBE) \cite{GL96,GPP96}, i.e. pseudoscalar mesons.
In the latter  model the hyperfine splitting in hadrons is
due to the short-range part of the Goldstone boson exchange
interaction between quarks, instead of the OGE interaction of
conventional models. The GBE interaction is flavor dependent
and its main merit is that it reproduces the correct ordering
of positive and negative parity states in all parts of the
considered spectrum. Moreover, the GBE interaction induces a
strong short-range repulsion in the $\Lambda$-$\Lambda$
system, which suggests that a deeply bound H-particle should
not exist \cite{ST98}, in agreement with the
high-sensitivity experiments at Brookhaven \cite{ST97}.

In the stability problem we are interested in the quantity
\begin{equation}
\Delta E =\ E(q^m\overline{q}^n) - \ E_T
\end{equation}
where $E(q^m\overline{q}^n)$ represents the multiquark energy and
$E_T$ is the lowest threshold energy
for dissociation
into two hadrons: two mesons for tetraquarks, a baryon + a meson
for pentaquarks and two baryons for hexaquarks.
In the right-hand side of (1) both terms are calculated within the same model.
A negative $\Delta E$
suggests the possibility of a stable compact mutiquark system.

We first give a brief summary of the situation for tetraquarks,
pentaquarks and hexaquarks. Next we outline the main characteristics
of the GBE model, which is more recent and less known than the OGE 
model. In the last section we compare the results which we obtained in the
frame of the GBE model with those from the literature, based on the
OGE model.

%%%%%%%%%%%%%%%%%%%%%%%%%%%%%%%%%%%%%%%%%%%%%%%%%%%%%%%%%%%%%%%%%%%%%%%
\section*{The Tetraquarks}
The {\it light} tetraquarks are related to the study of meson-meson scattering
as e.g. $\pi \pi$, $\pi \eta$, $\pi \eta'$, $\pi K$, etc.
and to the identification of scalar mesons,
i.e. mesons with quantum numbers
$J^{PC}$ = $0^{++}$, having masses and decay properties which do not fit 
into a $q \overline{q}$ bound state. 
One expects a non-$q \overline{q}$ scalar component to play an
important role in the mass range below 1800 MeV.
The well observed
isovector $a_0(980)$ and the isoscalar $f_0(980)$ mesons 
are interpreted as being $q^2 \overline{q}^2$ states \cite{JA77}
or in a more realistic version as $K \overline{K}$ molecules \cite{WI83}
(see also \cite{BS94}). 
It turns out that the $I = 0$ states are the most complex
both experimentally and theoretically \cite{PDG98}. 
Besides the $f_0(980)$
meson, there one has identified another three resonances namely 
$f_0(400-1200)$ or $\sigma$, $f_0(1370)$ and $f_0(1500)$. The  $f_0(1370)$
is thought to be a $u \overline{u} + d \overline{d}$ state
and the $f_0(1500)$ admits a glueball interpretation. The interpretation
of $\sigma$ remains open. A $q^2 \overline{q}^2$ state
is not excluded \cite{SCH99}. A recent reanalysis
of the $\pi \pi$ scattering \cite{ISHIDA99} reduces the interval
of the $\sigma$ mass to 400-800 MeV so that its central value
returns to 600 MeV. The scalars $a_0(980)$, the $f_0(980)$, the
$\sigma(600)$ and the $\kappa(900)$ (found in the analysis of
$\pi K$ scattering) are possible members of a scalar nonet
(see e.g. \cite{DE99}) and satisfy the Gell-Mann-Okubo
mass formula \cite{IS99}.

The {\it heavy} tetraquarks have been studied in a variety of models
and experimental search of double charmed tetraquarks $(cc\overline u
\overline d)$ are planned at CERN \cite{MO96}. In the
following let us denote by $q$ a light quark $u,d$ or $s$ and by $Q$
a heavy one $c, b$ or $t$. Theoretical work has focussed on tetraquarks
of type $QQ \overline q \overline q$ or equivalently $\overline Q
\overline Q qq$ (see e.g. \cite{BS98}). Note that one can also have  
$Q \overline Q q \overline q$ systems. These have two distinct
thresholds $Q \overline Q$ + $q \overline q$ and $Q \overline q$ +
$\overline Q q$. The latter is the same as for $\overline Q
\overline Q qq$ and it can be shown \cite{JMR94} that 
$m_{Q \overline Q} + m_{q \overline q} \leq 2 m_{Q \overline q}$ 
which means that $Q \overline Q$ + $q \overline q$ is the lower
threshold. Then assuming that the mass of $Q \overline Q q \overline q$
is the same as that of $QQ \overline q \overline q$, the latter has
more chance to be bound. 
Also, it is more convenient for variational studies, where an upper
bound is more conclusive about stability. Moreover $QQ \overline q
\overline q$ has no meson-antimeson annihilation channels as 
$Q \overline Q q \overline q$ does have. As an example, detailed
arguments in favour of the stability of $cc \overline u \overline d$
as compared to $c \overline c d \overline u$ are given in \cite{MO96}.

\begin{table} [h!]
\caption{Estimate of the size 
$R \sim (\alpha_s(m_Q)~ m_Q)^{-1}$ of a $QQ$ pair}
\label{table1}
\begin{tabular}{cccc}
QQ & $m_Q$(GeV) & $\alpha_s(m_Q)$ & R(fm) \\
\tableline
cc & 1.5 & 0.44 & 0.29 \\
bb & 5.0 & 0.28 & 0.14 \\
tt & 175.0 & 0.13 & $9.10^{-3}$ \\
\end{tabular}
\end{table}

The stability of $QQ \overline q \overline q$ relies on the fact that
QQ brings a small kinetic energy into the system and forms a tightly
bound pair of size $(\alpha_s(m_Q) m_Q)^{-1}$ (see Table
\ref{table1}). Then two 
heavy quarks act as an almost point-like heavy color antitriplet source.
If $Q$ is heavy enough, as it is the case of $t$,
the short range Coulomb attraction plays an important role in the 
formation of the tetraquark system and leads to $tt \overline q
\overline q$ stable states.
The claim in Ref. \cite{MW93} is that $c$ and $b$ are not heavy enough
to enter such a mechanism. The alternative is the existence of
weakly bound two heavy meson systems due to a potential determined
at long distance by one pion exchange and calculable in chiral
perturbation theory.
One pion exchange meson-meson interactions have also been discussed
in Refs. \cite{TO91} and \cite{EK93}.

Lattice gauge calculations became also recently available
\cite{MH99} and they may help in shedding more light
into the intermeson potential and to isolate contributions of
various mechanisms \cite{BBDS99}.

As mentioned above here we discuss results for stability looking
at the quantity (1). We compare value of $\Delta E$ obtained in the
literature from the OGE model \cite{BS93} with those obtained in
\cite{PSGR97} from the GBE model. Ref. \cite{PSGR97}
considers only the most favourable
configuration which is $\overline 3 3$ $S = 1, I = 0$. This means that
$QQ$ is in a $\overline 3$ color state and $\overline q \overline q$
in a 3 color state. The mixing of the $6 \overline 6$ is neglected
because one expects that this plays a negligible role in deeply
bound heavy systems. Then the Pauli principle rquires $S_{12} = 1$
for $QQ$ and $S_{34} = 0, I_{34} = 0$ for $\overline q \overline q$,
if the relative angular momenta are zero for both subsystems.
This gives a state of total spin $S = 1$ and isospin $I = 0$.
In the channel of light quarks having  $S^{light} = S_{34} = 0,
I^{light} =  I_{34} = 0$, as above, the lattice gauge calculations
\cite{MH99} produce a strong short-range attraction for $bb \overline q
\overline q$, which is consistent with constituent quark model
calculations \cite{BS98,BS93,PSGR97}.

%%%%%%%%%%%%%%%%%%%%%%%%%%%%%%%%%%%%%%%%%%%%%%%%%%%%%%%%%%%%%%%%%%%
\section*{The Hexaquarks}
We shall discuss hexaquarks before pentaquarks because they have
been proposed first in historical order \cite{JA77}.

In the {\it light} sector the well known example is the H-particle
which has been extensively studied in the literature. 
A recent and comprehensive review can be found in Ref. \cite{SSY}.
From the time it was proposed by Jaffe lots of theoretical studies
have been performed within a variety of models as the bag model,
the Skyrme model, constituent quark models, lattice calculations, 
QCD sum rules, etc. The results spread over a wide range of
predictions depending on the model parameters and the approximations
involved. In each model there are predictions for a bound
state or for an unstable state. In the flavor singlet $uuddss$ system 
with $J^P = 0^+, I = 0$ the GBE model induces a
strong repulsion of 847 MeV above the $\Lambda \Lambda$ threshold 
\cite{ST98}. This implies that the H-particle should not exist
as a compact object, in contrast to Jaffe's picture. A molecular 
type structure, as that of the deuteron, is not excluded however. 

In the {\it heavy} sector attention has been focused on hexaquarks
of type $uuddsQ$.  These systems are like the H-particle where one of the $s$
quarks has been replaced by a heavy one. Then when $Q = c$ the
particle is denoted by $H_c$ and when $Q = b$ the particle is denoted by
$H_b$. In the context of a diquark model \cite{LI97} the charmed
hexaquark is found to be unstable but the bottom hexaquark is 
found to be stable by about 10 MeV. Calculations based on a chromomagnetic
interaction give both $H_c$(I=0,J=3) and $H_b$(I=0,J=2 or 3) stable by
7.7 MeV up to 13.8 MeV \cite{LS93}. In the GBE model both $H_c$ 
and $H_b$ turn out to be unstable, with
about the same amount of repulsion above the respective thresholds,
as for the H-particle \cite{PS98}. This means that the heavy flavor
has no effect on the stability in these cases. 

%%%%%%%%%%%%%%%%%%%%%%%%%%%%%%%%%%%%%%%%%%%%%%%%%%%%%%%%%%%%%%%%%%%  
\section*{The Pentaquarks}

The pentaquarks $P_{\overline c s}^0$ = $uuds \overline c$
and $P_{\overline c s}^-$ = $udds \overline c$
have been proposed as stable systems against strong decays nearly
simultaneously in Refs. \cite{GI87,LI87}, about ten years after
Jaffe's proposal for the H-particle.
The more realistic calculations 
\cite{LB89}
which take into account 
the SU(3)-flavor
breaking, etc. also lead to stability. In  the OGE model the
stable pentaquarks have negative parity (i.e. the parity of the
antiquark) and require strangeness.

In the GBE model the best candidates to stability are not necessarily
strange and have positive parity \cite{STA98}. To understand these
differences let us make the simplifying assumption
that the heavy antiquark has an infinite mass. Then we need to care 
only about the light quarks, which we assume identical.
The wave function of the light subsystem, containing radial, spin,
flavor and color parts must be antisymmetric. The color part has necessarily
the symmetry $[211]_C$ in order to form a color singlet together with
the antiquark.
Let us consider the spin $S = 0$ state of $q^4$. There are two ways
to construct a totally antisymmetric state:\par
1) assume that the orbital part is symmetric
i. e. has symmetry $[4]_O$. Then the flavor-spin part must have
the symmetry $[31]_{FS}$. The inner product rules of the permutation
group \cite{FS96} 
require that the flavor part must be $[211]_F$. In the FS coupling
this state reads

\begin{equation}
\label{state1}
\left.{\left|{1}\right.}\right\rangle\ =\
\left({{\left[{4}\right]}_{O}{\left[{211}\right]}_{C}{\left[{211}\right]}_{
OC}\
;\
{\left[{211}\right]}_{F}{\left[{22}\right]}_{S}{\left[{31}\right]}_{FS}}\right)
\end{equation}

This state has $L=0$, thus positive parity. Together with the antiquark
this leads to a ${\it negative}$ parity pentaquark. Obviously 
its flavor part $[211]_F$ requires strangeness. In the GBE model
these pentaquarks are unbound \cite{GRSP98}.
In the CS coupling the state (\ref{state1}) has the 
same form but with C and F indices interchanged. It is the most favourable
state in the OGE model \cite{GI87,LI87}, because it has the lowest allowed
symmetry in the $CS$ space.

2) assume that the flavor-spin part is symmetric. Then the
Pauli principle requires that the orbital part should have the symmetry 
$[31]_O$. The spin state is $[22]_S$, as before, so that inner product
rules require the symmetry $[22]_F$ for the flavor part in order
to get $[4]_{FS}$.
In the FS coupling this state reads 
\begin{equation}
\label{state2}
\left.{\left|{2}\right.}\right\rangle\ =\
\left({{\left[{31}\right]}_{O}{\left[{211}\right]}_{C}{\left[{{1}^{4}}\right]}_{
OC}\
;\
{\left[{22}\right]}_{F}{\left[{22}\right]}_{S}{\left[{4}\right]}_{FS}}\right)
\end{equation}

The lowest angular momentum associated to $[31]_O$ is $L=1$ so that
this state has negative parity and together with the antiquark gives
a ${\it positive}$ parity pentaquark. The $[22]_F$ symmetry indicates that
strangeness is not required. The state $[4]_{FS}$ is the most favourable
in the GBE model because it has the lowest symmetry in the $FS$ space
and, as will be shown in the next section, the GBE hyperfine interaction has 
a flavor-spin operator which, of course, takes the lowest expectation value for 
the most symmetric $FS$ state.

The extension to the heavy flavor sectors of the Skyrmion approach
\cite{CK85} allowed to calculate the spectra of the lowest lying
pentaquarks containing charm and bottom antiquarks \cite{RS93}.
Interestingly, the conclusions are similar to those of the GBE model:
1) the lowest pentaquarks have positive parity for any flavor content
and 2) strangeness is not necessary in order to gain stability.

Finally, binding due to the long range one pion exchange has also
been considered \cite{SH95} and leads to a molecular type structure.

%%%%%%%%%%%%%%%%%%%%%%%%%%%%%%%%%%%%%%%%%%%%%%%%%%%%%%%%%%%%%%%%%%%
\section*{The GBE model}

Here we refer to the GBE model as originally
proposed by Glozman and Riska \cite{GL96}.
Its present status can be found in Ref. \cite{LYG99}.
Besides the pseudoscalar meson exchange, both the vector and scalar
meson exchanges are now incorporated in the model.
The calculations presented here are based on the nonrelativistic version
of Ref. \cite{GPP96}.

The origin of the model lies in the spontaneous breaking of chiral symmetry in QCD
which implies the existence of constituent quarks with a dynamical
mass and Goldstone bosons (pseudoscalar mesons).
Accordingly, it is assumed that the underlying dynamics in the low
energy regime is due to Goldstone boson exchange between constituent
quarks. In a nonrelativistic reduction for the quark spinors
the quark meson vertex is proportional to
$\vec{\sigma} \cdot \vec{q}~ \lambda^F$, with 
$\lambda^F$ the Gell-Mann matrices, $\vec{\sigma}$ the Pauli matrices,
and $\vec{q}$ the momentum of the meson. This generates a meson exchange
interaction which is spin and flavor dependent. In the coordinate space the
corresponding interaction potential contains two terms. One is the Yukawa
potential tail and the other is a contact $\delta$-interaction.
When regularized,  this generates the short range part of the 
quark-quark interaction.
It is this short range part which dominates over the Yukawa part
in the description of baryon spectra and leads to a corect order
of positive and negative parity  states both in nonstrange and
strange baryons. 

The dominant interaction is reinforced by the short-range part of
the vector meson exchange (two correlated pions) \cite{RB99}.

The model is supported by the independent phenomenological analysis
of the $L = 1$ baryons \cite{CG99},
by the $1/N_c$ expansion studies of the $L = 1$ nonstrange baryons
\cite{CC99} and by lattice studies \cite{LIU99}.

\begin{table}[h!]
\caption{Results for $\Delta E$ for charmed tetraquarks, pentaquarks
and hexaquarks. Each case corresponds to the most favourable
$I,J^P$ state.}
\label{table2}
%\begin{tabular}{lrrr}
\begin{tabular}{lccc}
%\begin{tabular}{lddd}
System& $Parity$&
   \multicolumn{1}{c}{$\Delta E$ for OGE\tablenote{In all cases a nonrelativistic
Hamiltonian is used. It contains a linear confinement
and a chromomagnetic spin-spin interaction
.}} &
  \multicolumn{1}{c}{$\Delta E$ for GBE\tablenote{We use the nonrelativistic
version of Ref. \cite{GPP96}}}\\
\tableline
$uu \overline c \overline c$ & + & 19 MeV (I=0,J=1)\cite{BS93} & $-$185
Mev (I=0,J=1) \cite{PSGR97}. \\
$uuds \overline c$ & - & -51 MeV (I=$\onehalf$,J=$\onehalf$) \cite{LB89} &
488 MeV (I=$\onehalf$,J=$\onehalf$) \cite{GRSP98}\\
$uudd \overline c$ & + & unbound & $-75.6$ MeV (I=0,J=$\onehalf$)
\cite{STA98}\\
$uuddsc$ & + & -7.7 MeV (I=0,J=3) \cite{LS93}
 & 625 MeV (I=0,J=0 or 1) \cite{PS98} \\
\end{tabular}
\end{table}

\subsection*{A schematic GBE model}
In a schematic model the dominant GBE interaction takes the form
\begin{equation}
\label{VCHI}
{V}_{\chi }\ =\ -\ {C}_{\chi }\ \sum\limits_{ i\ <\ j}^{} {\lambda }_{
i}^{ F} \cdot {\lambda }_{ j}^{ F} \ {\vec{\sigma }}_{i} \cdot {\vec{\sigma
}}_{j}
\end{equation}
with $C_{\chi}\cong$ 30 MeV, determined from the $\Delta$-N splitting 
\cite{GL96}. It is useful to give an estimate of $\Delta E$, based on 
the interaction (\ref{VCHI}). As an example, let us consider 
pentaquarks containing heavy flavor, for which equation (1) becomes

\begin{equation}
\label{PENTA}
\Delta E =\ E(q^4\overline{Q}) - E(q^3) - E(q \overline{Q})
\end{equation}

First we suppose that the confinement energy roughly cancels out in
(\ref{PENTA}). Next, as in the previous section, we suppose that
$m_Q \rightarrow \infty$. As a consequence, the quark-antiquark interaction
can be neglected in the expectation value of (\ref{VCHI}) both for
the pentaquark and the emitted heavy meson. Using the Casimir operator
technique one finds that the contribution of $V_\chi$ to $E(q^3)$
is $-14~C_{\chi}$. Now we have to distinguish the two cases introduced
in the previous section.

\paragraph*{Negative parity pentaquarks}. In this case, in a harmonic
oscillator basis, the contribution of the kinetic energy 
to $\Delta E$ is $\threequarters~\hbar \omega$. This difference is
exactly the kinetic energy associated to the extra degree of freedom
in the pentaquark, corresponding to the relative motion between the
$q^3$ and $q \overline Q$ subsystems. Using again the Casimir operator
technique one finds that the state (\ref{state1}) leads to 
$\langle V_{\chi} \rangle = - 16 C_{\chi}$. The separation energy
becomes

\begin{equation}
\label{negative}
\Delta E = \threequarters~\hbar \omega - (16 - 14)~C_{\chi} = 128~ MeV.
\end{equation}
where the numerical value  results from taking $\hbar \omega = 250$ MeV
and $C_{\chi} = 30$ MeV \cite{GL96}. 
The fact that $\Delta E$ is positive indicates that the GBE interaction
leads to unbound negative parity pentaquarks of a compact type. This is 
confirmed by the more precise estimates \cite{GRSP98} where $\Delta E$
obtained from a variational method is several hundred MeV for all
strange or nonstrange pentaqurks under consideration, 
containing $c$ or $b$ antiquarks.

\paragraph*{Positive parity pentaquarks}. In this case the state
(\ref{state2}) suggests that there is a unit of orbital excitation in
the pentaquark due to the symmetry state $[31]_O$ of the $q^4$
subsystem. This leads to $\Delta K.E. = 5/4 ~\hbar \omega$. But at the
same time the contribution of the spin-flavor interaction becomes
more attractive than for negative parity pentaquarks, giving 
$\langle V_{\chi} \rangle = - 28 C_{\chi}$ due to the higher symmetry
$[4]_{FS}$ present in (\ref{state2}). Then one has
 
\begin{equation}
\label{positive}
\Delta E = 5/4~\hbar \omega - (28 - 14)~C_{\chi} = -100~ MeV.
\end{equation}
 
This proves that 
the attraction due to $ V_{\chi}$ overcomes the excess in the kinetic
energy due to the orbital excitation. This cannot happen for the OGE
interaction which has a spin-color structure, thus is flavor-blind,
and does not distinguish between the $[31]_{FS}$ and the $[4]_{FS}$
states. For this reason the positive parity pentaquarks are expected
to be even more unbound than the negative parity ones. In particular
the $uudd \overline c$ pentaquark will be unbound in any OGE model
(see Table \ref{table2})  inasmuch as the OGE interaction predicts
unbound negative parity pentaquarks of the same flavor \cite{LB89}.

%%%%%%%%%%%%%%%%%%%%%%%%%%%%%%%%%%%%%%%%%%%%%%%%%%%%%%%%%%%%%%%%%%%
\section*{Discussion}

The main objective of this talk is to compare results for stability
obtained from two constituent quark models: OGE and GBE.
We illustrate the discussion with results for charmed multiquark
systems as shown in Table
\ref{table2}.
From this table and the results reported above for the H-particle one
important conclusion can be drawn: when the GBE interaction stabilizes a 
system, the OGE interaction destabilizes it and vice-versa.

When the quark $b$ is used instead of $c$ the results are not
so strickingly different but still show large differences in the
predictions of the two models.

The challenging
question of the existence of exotics remains unanswered so far.
It is worthwhile to perform more elaborate calculations, based for
example, on the resonating group method, in order to better understand
the role played by various mechanisms and the dynamics
of exotics. The experimental search would be of great help in putting
constraints on the various effective quark-quark interactions.
\vspace{0.5cm}

\paragraph*{Acknowledgements.} I am most grateful to the organizing committee
and in particular to Professor J. da Providencia
for kindly inviting me to give this talk.

%%%%%%%%%%%%%%%%%%%%%%%%%%%%%%%%%%%%%%%%%%%%%%%%%%%%%%%%%%%%%%%%%%%%

\end{document}